# High Yield Growth and Doping of Black Phosphorus with Tunable Electronic Properties

Mingqiang Liu[1+], Simin Feng[1+], Yi Hou[1], Shilong Zhao[2,1], Lei Tang[1], Jiaman Liu[1], Feng Wang[2,1], Bilu Liu[1]*

[1] Shenzhen Geim Graphene Center (SGC), Tsinghua-Berkeley Shenzhen Institute (TBSI), Tsinghua University, Shenzhen 518055, P. R. China.
[2] Department of Physics, University of California, Berkeley, Berkeley, CA 94720, USA

Correspondence should be addressed to B.L. bilu.liu@sz.tsinghua.edu.cn
[+]M. Liu and S. Feng contributed equally.

**Keywords:** black phosphorus, doping, short-distance transport growth, electronic structures, stability

## Abstract

**Black phosphorus (BP) has recently attracted significant interest due to its unique electronic and optical properties. Doping is an effective strategy to tune a material's electronic structures, however, the direct and controllable growth of BP with a high yield and its doping remain a great challenge. Here we report an efficient short-distance transport (SDT) growth approach and achieve the controlled growth of high quality BP with the highest yield so far, where 98% of the red phosphorus is converted to BP. The doping of BP by As, Sb, Bi, Se and Te are also achieved by this SDT growth approach. Spectroscopic results show that doping systematically changes its electronic structures including band gap, work function, and energy band position. As a result, we have found that the air-stability of doped BP samples (Sb and Te-doped BP) improves compared with pristine BP, due to the downshift of the conduction band minimum with doping. This work develops a new method to grow BP and doped BP with tunable electronic structures and improved stability, and should extend the uses of these class of materials in various areas.**

Doping can significantly change the electronic, optical, magnetic, and chemical properties of materials, and has been widely used in semiconducting electronic and optoelectronic devices. The silicon semiconductor industry uses doping to change the transport characteristics of silicon-based electronic devices. In recent years, two-dimensional (2D) layer materials, such as graphene and transition metal dichalcogenides (TMDCs) have attracted much attention due to their unique properties and potential use in next-generation electronics, optoelectronics, etc[1-3]. In the context of 2D materials, doping is also useful for changing their properties, such as to produce p- or n-type transport behavior, and changing Fermi levels, and bandgaps. Theoretical and experimental studies show that substitutional doping changes the band structure and opens a bandgap in graphene[4, 5], which significantly expands the applications of graphene. For instance, Wei et al. have grown N-doped graphene by a chemical vapor deposition (CVD) method and demonstrated that N atoms were substitutionally doped in the graphene lattice to produce n-type behavior, indicating that substitutional doping has changed the electrical properties of graphene[5]. Dai et al. reported the N-doping of graphene through the electrothermal reaction of graphene with ammonia, suggesting that edge doping was a new approach to dope graphene ribbons that affected its properties[6]. In addition, Lv et al. have achieved substitutional doping of graphene with different atoms (B, N, or Si), and shown the use of doped graphene for molecule and gas sensing[7-9]. Besides graphene, Gong et al. have achieved substitutional doping of molybdenum disulfide ($MoS_2$) by selenium with a broad range of selenium concentrations, resulting in changes in the optical bandgaps[10]. It is therefore clear that doping is an effective strategy to change the properties of 2D materials.

Black phosphorus (BP) is a newly emerging 2D material which has a high charge carrier mobility,

a tunable band gap from visible to the mid-infrared region, and anisotropic electronic and optical properties, making it promising in high performance electronics and optoelectronics especially in the infrared range[11-15]. To realize BP-based electronics and optoelectronics, being able to change its electronic properties is of great importance and doping could be an effective strategy to achieve this. For instance, Kim et al. reported being able to change the bandgap of few-layer BP over a wide range by doping with potassium using an in situ surface doping technique[16]. However, compared to graphene and TMDCs, the direct growth and controlled doping of BP is still a challenge and less developed. Currently, BP is mostly grown by four methods: high temperature high pressure (HTHP) conversion of red phosphorus (RP)[17], bismuth flux[18-20], mercury flux[21], and chemical vapor transport (CVT) growth[22-24]. Among these methods, HTHP needs a very high pressure, up to a few GPa, which it is not feasible to generate economically. The bismuth and mercury flux methods are usually time consuming and use toxic chemicals, and have the shortcomings of a low conversion percentage of BP and small crystal sizes. Compared with other methods, CVT can grow BP at moderate temperature and pressure, and has recently been widely used to grow BP and doped BP[16, 25-29]. So far, the growth yield of BP (e.g., percentage conversion from RP or white phosphorus to BP) is usually low, and the products are a mixture of BP with other impurities like $Au_3SnP_7$, AuSn, $Sn_4P_3$, and $Sn_3P_4$. Although the doping of BP has been reported by few pioneering studies, the doping concentrations are usually low, e.g., 0.1 at% in Te-doped BP[29], and 0.39 at% in Se-doped BP[27]. In addition, it is not clear how such doping affects the electronic structure and properties. Therefore, it is critical to develop a new approach to achieve a high yield growth and a high doping concentration to systematically tune its properties, which is an important prerequisite for its use.

Here, we report an efficient short-distance transport (SDT) growth approach and achieve a growth of 98% BP from RP, the highest value reported so far. This SDT approach can also directly grow BP doped with various elements like As (As-BP or b-$As_xP_{1-x}$ with x in the range of 0 to 0.72), Sb (Sb-BP), Bi (Bi-BP), Se (Se-BP), and Te (Te-BP). Note that elements such as Sb, Bi, Se, and Te that are normally difficult incorporate into BP, have been doped by our SDT growth method with doping concentrations as high as 1.2, 0.36, 0.45 and 0.42 at%, respectively, and these are among the highest doping concentrations reported so far. Significantly, materials characterization shows that the doping of BP effectively changes its electronic structures including bandgap and work function, which leads to an enhanced air stability of doped BP flakes compared with pristine BP flakes. Atomic force microscopy (AFM) has been applied to analyze the topographic features of both pristine and Sb and Te doped BP flakes as a function of ambient exposure time and we observed that doped BP flakes show a much smaller surface roughness increase than pristine BP flakes. Our studies indicate that doping of a proper element is a promising route to suppress the ambient degradation of BP, and it will accelerate the implementation of BP in future applications.

BP is grown through a direct reaction of source materials followed by a transport agent-assisted SDT process in an evacuated quartz ampule at high temperatures (**Figure 1a**). Different from the mineralizer-assisted CVT growth method developed by Nilges et al. where a temperature gradient is needed in a long-distance transport (LDT) growth fashion (**Figure 1b**)[22, 23], our SDT growth approach uses a uniform temperature. The experimental details of this method are described in the Experimental Section. **Figure 1c,d** show photos of the ampoules after typical SDT and LDT growth, respectively. One can clearly see that a large quantity of BP is grown by the SDT process while only few BP crystals could be observed using the LDT process, clearly showing that SDT method produces a much higher BP yield than the LDT method. In order to understand the growth mechanism of BP in this SDT process, we have performed systematic experiments with different growth parameters, including the type of transport agent, the mass ratio of Sn/SnI$_4$, and the growth temperature. In order to study whether Sn and SnI$_4$ are necessary reactants during the growth of BP, we have used other iodides, such as PbI$_2$, BiI$_3$, I$_2$, NH$_4$I, NaI and KI instead of SnI$_4$ as the reaction precursors to try to grow BP in both the SDT and LDT methods (**Figure S1, Supporting Information**). We found that Sn is indispensable for BP growth, and no BP will grow without the Sn catalyst, consistent with a previous report[30]. It should be noted that the growth yield of BP (conversion ratio of RP to BP, i.e., $m_{BP}/m_{RP}$) in the SDT method is always higher than in the LDT method (**Figure S4a, Supporting Information**). Furthermore, we have used different Sn/SnI$_4$ mass ratios (1:1, 1:2, 1:3, 1:4, 2:1, 2:3, 3:1 and 3:2), **Figure S2 (Supporting Information)** shows photographs of BP grown under different conditions. **Figure S4b (Supporting Information)** compares the BP growth yield using the two methods with different Sn/SnI$_4$ mass ratios. Based on this series of experiments, we conclude that the highest BP growth yield is obtained when the Sn/SnI$_4$ mass ratio is 2:1. The growth yield of BP at this condition is over 98%, which is the highest value reported so far with previously reported results in the range 50 to 90%[22-24, 31-33]. In addition, to understand the effect of temperature on BP growth, we have carried out a series of experiments at different temperatures and the growth yield of BP is shown in **Figure S3** and **Figure S4c (Supporting Information)**. These results show that BP crystals are grown at temperatures ranging from 400 to 800 °C, while the growth yield of BP is a maximum when the temperature is between 600 °C and 750 °C. This result is understandable because a low temperature would decrease the growth rate of BP, while a high temperature would make its nucleation difficult[34, 35]. After growth, the purity of the BP was characterized by energy dispersive X-ray spectroscopy (EDX) analysis (**Figure S6, Supporting Information**). **Figure 1e** summarizes the growth yield of BP using the different methods reported in the literature (HTHP method[28, 29, 36, 37], sonochemistry[38], flux method[19, 39, 40] and CVT method[22, 23, 31-33, 41]). Overall, the results show that a high growth yield and high purity BP are obtained by the SDT method developed in this work using a uniform temperature.

With these control experiments, we conclude that a uniform temperature and short-distance transport are two critical factors in achieving a high BP growth yield. When the temperature of the quartz ampoule slowly increases, RP, Sn, and SnI$_4$ reactants begin to sublimate and fill the ampoule. Phosphorus gas reacts with Sn and SnI$_4$ gas to form P-Sn-I compounds, which condense along the bottom of the ampoule, serving as nucleation sites for subsequent BP growth. During the slow cooling, the crystallization and growth of BP is assisted by the transport agent until the phosphorus gas is completely converted into BP. This process and growth mechanism is shown in **Figure S5 (**see details in the Supporting Information**)**.

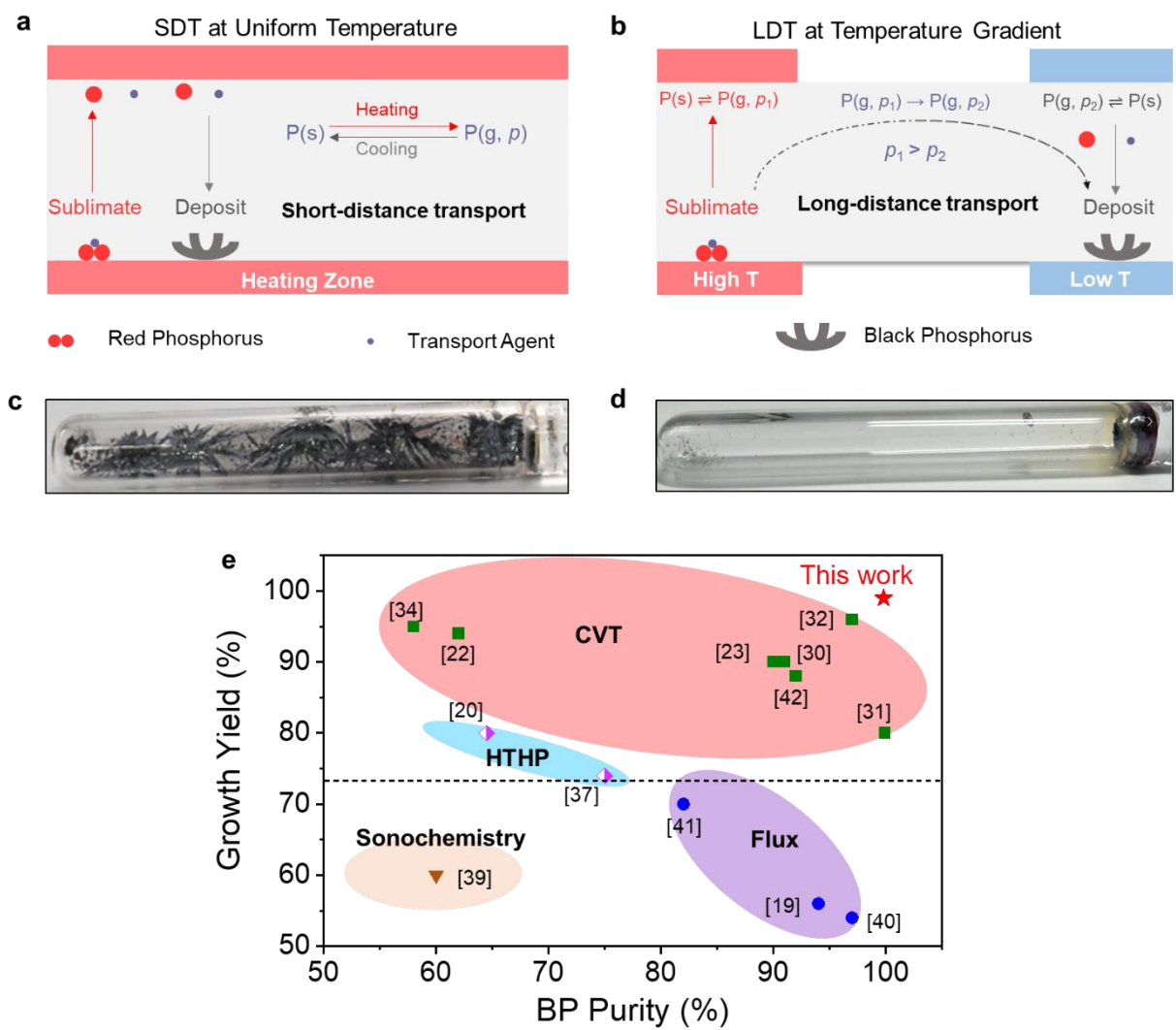

**Figure 1. SDT growth of BP under a uniform temperature with the highest growth yield and purity.** (a), (b) Schemes of uniform temperature SDT and temperature gradient LDT strategy to grow BP. (c) and (d) Photographs of the as-grown products in the SDT and LDT growth methods. (e) A summary of literature reports on the growth yield and purity of BP grown by various methods.

We have used a comprehensive suite of microscopy and spectroscopy techniques to characterize the as-grown BP. **Figure 2a** shows a photograph of BP crystal laths, densely distributed in a quartz ampoule and naturally separated from any by-products after growth. Optical (OM) and scanning electron microscope (SEM) images clearly show that the BP crystal has a lath-like morphology (**Figure 2b** and **Figure 2c**). High resolution transmission electron microscopy (HRTEM) studies were performed after exfoliating the BP onto a TEM grid. In **Figure 2d,e**, a representative HRTEM image and the corresponding selected-area electron diffraction (SAED) pattern provide evidence that the BP flakes have an orthorhombic crystal structure. Lattice fringes with a 0.32 nm separation correspond to the (100) plane of BP, which is in good agreement with a previous report[42]. EDX spectroscopy results show that our BP samples are of high purity (**Figure S6, Supporting Information**). X-ray photoelectron spectroscopy (XPS, **Figure 2f**) indicates that the P $2p_{1/2}$ and $2p_{3/2}$ core level peaks are located at 130.8 and 129.9 eV, respectively, in good agreement with previously reported results for BP[22-24, 31-33]. The absence of peaks from the oxidation states of phosphorous ($PO_x$) at 135.0 eV suggests that the as-grown BP is not oxidized.

The crystal structure of BP was also examined by X-ray diffraction (XRD, **Figure 2g**). All the diffraction peaks can be indexed to BP crystals with a preferred orientation of *(0k0)*. The most intense peak in the XRD pattern (**Figure 2h)** is the (040) peak which has a narrow full width at half maximum (FWHM) of 0.186°, indicating good crystallinity[39]. This was supported by Raman and photoluminescence (PL) spectroscopy results. The Raman spectrum (**Figure 2i**) shows three distinct vibration modes at 360.9, 437.6 and 465.4 cm$^{-1}$, respectively corresponding to the $A_g^1$, $B_{2g}$ and $A_g^2$ phonon modes of BP. The PL spectrum (**Figure 2j**) of monolayer BP, which was exfoliated from bulk BP and encapsulated by h-BN, shows a peak at 730.5 nm with a FWHM of 22.6 nm. This narrow FWHM correlates with the XRD results and supports the high crystallinity of BP crystals[43]. In addition, we have studied the optical absorption of monolayer BP by reflectance measurements at 77 K using a microscopy set-up (**Figure S7d, Supporting Information**). The result shows that monolayer BP has a bandgap of around 729 nm (1.70 eV, **Figure S7c, Supporting Information**), consistent with a previous report[12]. All these results imply that we have obtained high quality BP by the SDT growth method with a uniform temperature.

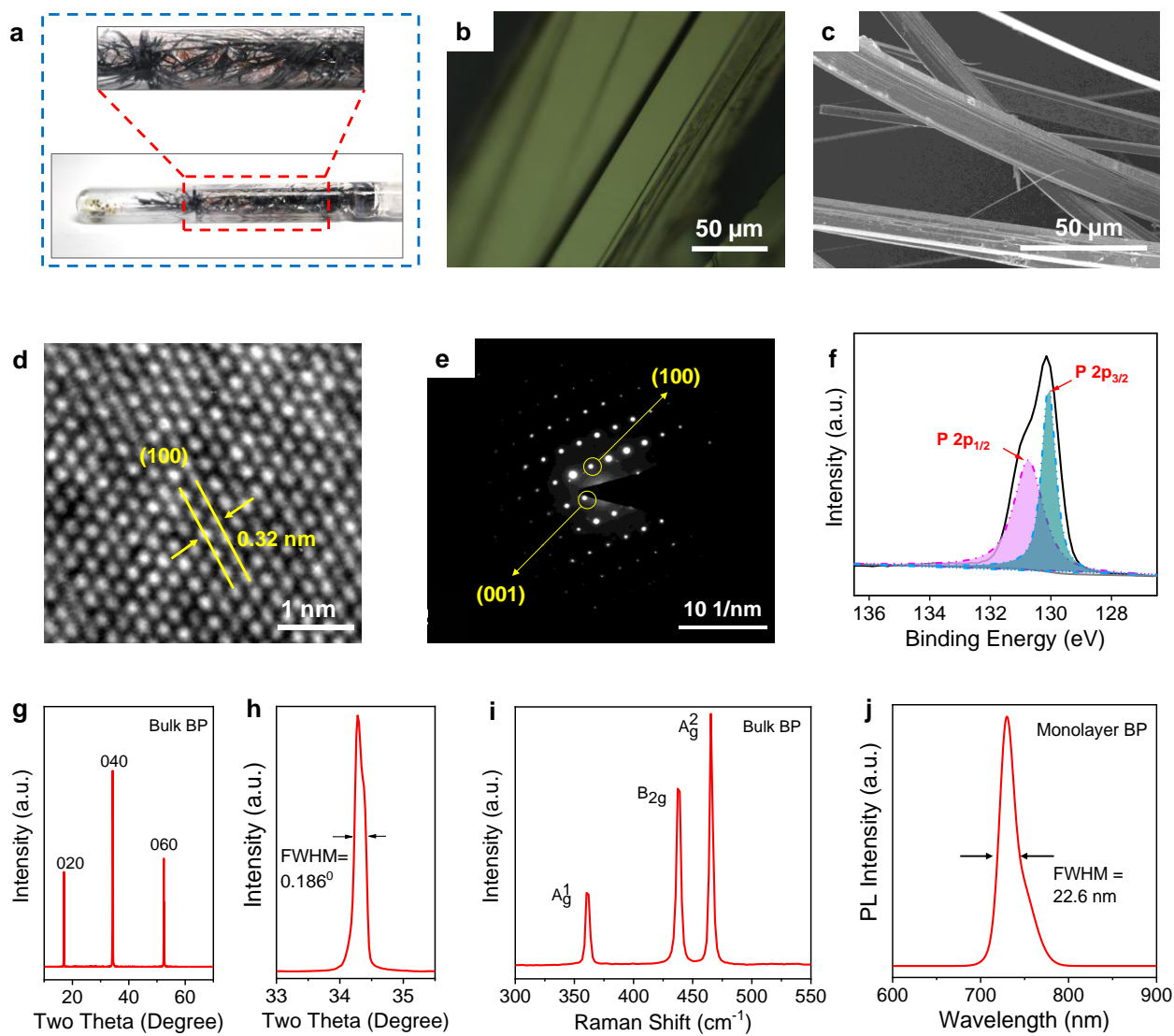

**Figure 2. Characterization of BP grown by the SDT method under uniform temperature.** (a) Photographs of the products in a quartz ampoule after the reaction. (b), (c) OM and SEM images of BP crystal laths. (d) HRTEM image of a BP flake showing its high crystallinity. (e) An SAED pattern of the BP sample. (f) XPS of P 2p spectrum. (g) XRD pattern of BP crystal. (h) Magnified section of the (040) peak in the XRD pattern with a FWHM of 0.186°. (i) Raman spectrum of a BP crystal. (j) PL spectrum of monolayer BP under 77 K in vacuum with the sample was encapsulated by h-BN, showing a PL peak at 730.5 nm with a FWHM of 22.6 nm.

We have used the same SDT growth method with a uniform temperature to grow BP doped with various elements, including As, Sb, Bi, Se, and Te. Details of the growth conditions are provided in the Methods section. **Figure S8** and **Figure S9** (**Supporting Information**) shows photographs of the products (Te-BP, Se-BP, Sb-BP, Bi-BP and b-$As_xP_{1-x}$) after SDT growth. They show that doped BP samples have been grown and are densely distributed in the quartz ampoules, showing a similar morphology and layer structure to the BP (**Figure S10, Supporting Information**). **Figure 3a** are the XRD patterns of doped BP samples compared with pristine BP and show a downshift of diffraction angles for doped BP, indicating that the crystal lattice expands after doping (**Figure 3b**). This result indicates an expansion of the BP crystal lattice as a result of doping and is attributed to the larger radius than P of the doping elements. We have also observed that with increasing atomic number of the doping elements (comparing Se with Te and Sb with Bi), the growth yield of the doped BP decreases slightly. Similarly, when the doping concentration increases, for example when the As dopant concentration increases from 0% (pure BP) to 72.3%, the growth yield also decreases (**Figure S9, Supporting Information** and **Figure 3c**). It should be noted that it has always been a challenge to achieve high quality BP growth with a high yield, especially when doped with other elements, but here the growth yield was always above 90%, indicating the effectiveness of this method in growing BP and doped BP.

In order to study the structural change in the BP during doping, we have used several characterization methods to demonstrate the successful doping of elements such as Te-, Se-, Sb- and Bi- in BP, which are usually difficult to incorporate. Raman analysis show that the $A_g^1$ (359.3 cm$^{-1}$), $B_{2g}$ (435.4 cm$^{-1}$), and $A_g^2$ (463.3 cm$^{-1}$) peaks slightly red shifted (by 1.6 cm$^{-1}$ for $A_g^1$, 2.3 cm$^{-1}$ for $B_{2g}$, and 2.1 cm$^{-1}$ for $A_g^2$), when BP was doped with other elements (**Figure 3d** and **Figure S11a, Supporting Information**). More interesting was the fact that two new peaks at 193.5 and 229.0 cm$^{-1}$ appeared for all the doped BP samples (**Figure S11b, Supporting Information**), which were not observed in BP. Previous theoretical studies suggested that these two peaks belong to the edge phonon vibrational modes of pristine BP[44, 45], which are identified as the $B_{1g}$ and $B_{3g}^1$ modes and are forbidden in pristine BP. Our results suggest that these two peaks may be activated by doping and could possibly provide a way to identify doped BP. The doping of the as-grown BP was directly proved by XPS measurements. Taking Sb-doped BP as an example, we have observed two strong peaks at 130.9 and 130.0 eV (**Figure 3e**), corresponding to the P $2p_{1/2}$ and P $2p_{3/2}$ peaks. In addition, two new peaks located at 538.3 and 528.9 eV are seen (**Figure 3f**), corresponding to the Sb $3d_{3/2}$ and Sb $3d_{5/2}$ peaks and indicating the successful doping of Sb in BP. In addition, we have performed HRTEM analysis, coupled with EDX mapping to investigate the microstructure and chemical composition of the doped BP samples (**Figure S12, Supporting Information**). The HRTEM images prove the good crystallinity of the doped BP samples and EDX mapping shows that Sb and Se dopants are uniformly distributed in the doped BP. Furthermore, we have performed inductively coupled plasma optical emission spectroscopy (ICP-OES) to quantitatively measure the dopant concentration and the results show that Sb-, Se-, Te- and Bi-doped BP contain dopant concentrations of 1.2, 0.45, 0.42 and 0.36 at%, respectively (**Figure 3g**). We found that it becomes more difficult to dope elements into BP as their atomic radius increases. It should be noted that the dopant concentrations for the different elements are among the highest ever obtained[27, 29], confirming the effectiveness of this SDT method to incorporate different elements into the BP lattice.

More interestingly, we have observed that arsenic (As), which is from the same chemical group as phosphorus, can be doped into BP crystals with high and defined concentrations in the uniform temperature SDT process. Raman characterization indicates that the intensity of the BP peaks decreases while the intensity of the peaks from b-AsP increases with increasing As concentration (**Figure 3h**). For the b-As$_{0.19}$P$_{0.81}$ sample, we have observed two strong peaks at 131.2 and 130.4 eV in XPS characterization (**Figure 3i**), corresponding to the P $2p_{1/2}$ and P $2P_{3/2}$ peaks. In addition, two new peaks at 41.8 and 42.4 eV related to As $3d_{5/2}$ and $3d_{3/2}$ appeared (**Figure 3j**), proving the existence of As in the sample. Various dopant concentrations from 0% (pure BP) to 72.3% (b-As$_{0.72}$P$_{0.28}$) are achieved by changing the ratios of the starting materials including RP and grey arsenic (**Figure 3k**), which is consistent with a previous report using LDT process[26, 46]. These results are consistent with previous results[26, 46], confirming the successful doping of As into BP.

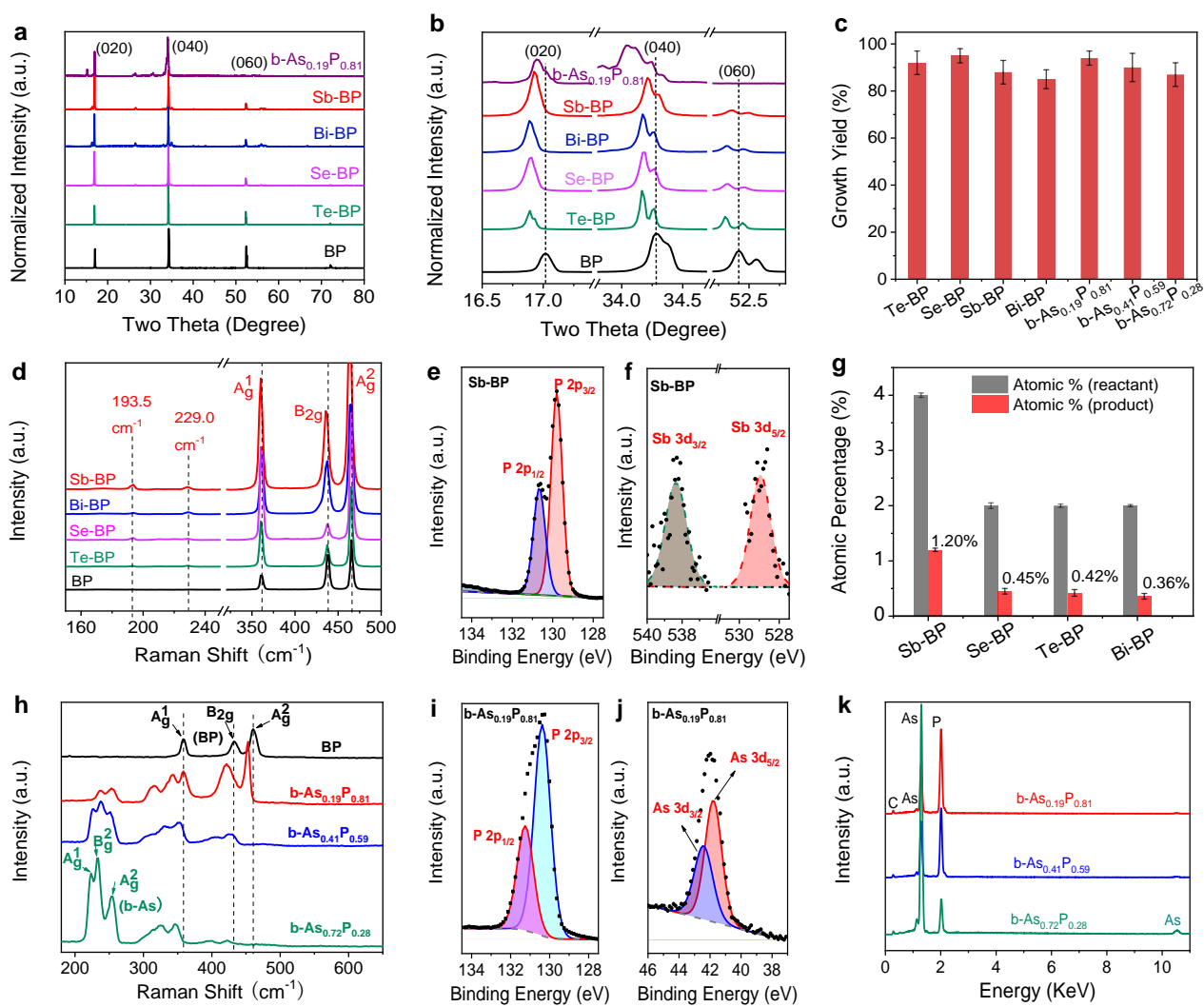

**Figure 3. Characterization of BP doped with various elements, such as Te, Se, Bi, Sb, and As.** (a) XRD patterns of BP, Te-BP, Se-BP, Bi-BP, Sb-BP and b-As$_{0.19}$P$_{0.81}$. (b) Expanded patterns in (a) from 16.5° to 55° showing details of the (020), (040) and (060) peaks. (c) Growth yields of Te-BP, Se-BP, Sb-BP, Bi-BP, b-As$_{0.19}$P$_{0.81}$, b-As$_{0.41}$P$_{0.59}$, and b-As$_{0.72}$P$_{0.28}$. (d) Raman spectra of Sb-BP, Se-BP, Te-BP and Bi-BP. (e) P 2p and (f) Sb 3d XPS spectra of Sb-BP. (g) ICP-OES results of Sb-BP, Se-BP, Te-BP and Bi-BP, showing the concentrations of each doping element. (h) Raman spectra of BP, b-As$_{0.19}$P$_{0.81}$, b-As$_{0.41}$P$_{0.59}$ and b-As$_{0.72}$P$_{0.28}$. The b-AsP Raman peaks gradually show up as As is doped into the BP lattice, while the BP Raman peaks fade away. (i) P 2p and (j) As 3d XPS spectra of b-As$_{0.19}$P$_{0.81}$. (k) EDX spectra of b-As$_{0.19}$P$_{0.81}$, b-As$_{0.41}$P$_{0.59}$ and b-As$_{0.72}$P$_{0.28}$.

We now focus on how doping changes the electronic structures of BP. Absorption spectra of As-, Sb-, Se-, Te-, and Bi-doped BP powder (the mixture of 5 mg sample powder and 100 mg KBr were grinding in a glove box filled with Ar gas) were examined using diffuse reflectance infrared Fourier transform (DRIFT) spectroscopy under ambient conditions (**Figure 4a**). The results show that after doping with these different elements the absorption band edges shift to the long wavelength region. The band gaps ($E_g$) of different samples can be calculated using the Tauc plot method (see Supporting Information)[45]. The ordinate of the spectra was calculated according to the Kubelka−Munk (K-M) method (see Supporting Information). Following this procedure, $E_g$ values of pristine BP, Se-BP, Sb-BP, Te-BP, Bi-BP, b-$As_{0.19}P_{0.81}$, b-$As_{0.41}P_{0.59}$, and b-$As_{0.72}P_{0.28}$ samples were obtained from the [$(h\nu*F(R_\infty))^2$ *vs.* h$\nu$] plots (**Figure 4b**), and calculated to be 0.284, 0.278, 0.283, 0.287, 0.274, 0.203, 0.187 and 0.168 eV, respectively (**Table S1, Supporting Information**). The bandgap of pristine bulk BP obtained here is in agreement with earlier measurements[12]. The results show that the bandgaps of BP are modulated by doping with Te doping causing an increase and As, Sb, Bi and Se doping a decrease. Interestingly, one can see from **Table S1 (Supporting Information)** that the band gap of b-$As_{0.19}P_{0.81}$ decreases sharply when the concentration of As increases from 0% (pure BP) to 19.3%, and then decrease less significantly when the concentration of As increases from 19.3% to 41.0% and further to 72.0%. The results suggest that the band gaps do not scale linearly with the concentration of arsenic or phosphorus. The band gap of b-$As_xP_{1-x}$ (x= 0.19, 0.41, and 0.72) obtained from infrared absorption measurements agrees with results from our previous work[26]. These results clearly show that the bandgap of BP is modulated by doping.

In addition to the bandgap, the band structure of BP is also tuned by doping. We have performed ultraviolet photoelectron spectroscopy (UPS, h$\nu$ = 21.22 eV, **Figure 4c-d**) measurements to study the band position of pristine BP and doped BP samples in high vacuum (ca. $10^{-8}$ Torr) without exposure to air, and these show the secondary electron cut-off energy ($E_{cutoff}$, **Figure 4c**) and the Fermi level ($E_{Fermi}$ =0.0 eV) and valence band maximum ($E_{VBM}$, **Figure 4d**) position of the samples. UPS was used to determine the work function ϕ (ϕ = hν- ($E_{cutoff}$ –$E_{F(Au)}$)), where $E_{cutoff}$ –$E_{F(Au)}$ denotes the secondary electron edge), and the position of the valence band maximum ($E_{VBM}$) relative to the Fermi level ($E_{Fermi}$). We calculated the $E_{VBM}$ - $E_{Fermi}$ of the pristine BP, Se-BP, Sb-BP, Te-BP, Bi-BP, b-$As_{0.19}P_{0.81}$, b-$As_{0.41}P_{0.59}$, and b-$As_{0.72}P_{0.28}$ samples to be 0.19, 0.14, 0.24, 0.26, 0.20, 0.10, 0.08 and 0.07 eV, respectively (**Table S1, Supporting Information**). As doped BP is As-doped, the $E_{Fermi}$ is expected to shift closer to the $E_{VBM}$ as filled states are emptied by electron transfer to the dopants, and the negatively charged doping products on the surface is expected to increase the vacuum level ($E_{vac}$), which can effects contributing to an increased work function (**Table S1, Supporting Information**). In particular, b-$As_{0.72}P_{0.28}$ shows a more remarkable upshift of the $E_{VBM}$ due to the high doping concentration of As. The evolution of the UPS spectra in the low kinetic energy region indicates that the work function of BP is remarkably modulated by doping (**Table S1, Supporting Information**), from 4.11 eV in BP to 4.25 eV in Sb-doped BP, and from 4.11 eV in BP to 4.21 eV in Te-doped BP. These results suggest that by controlled doping, a desired work function could be obtained in BP, and this is important for the electrical contact engineering of BP electronics and optoelectronics. In addition, by combining the $E_{VBM}$ with the bandgap, which can adjust the position of the conduction band minima ($E_{CBM}$) of the BP sample, and to improve air-stability of BP. These band alignment results (**Figure 4e**) are a useful guide for building BP-based heterojunctions and devices, because precise control of the band structure and alignment could engineer the carrier transport efficiency of the system.

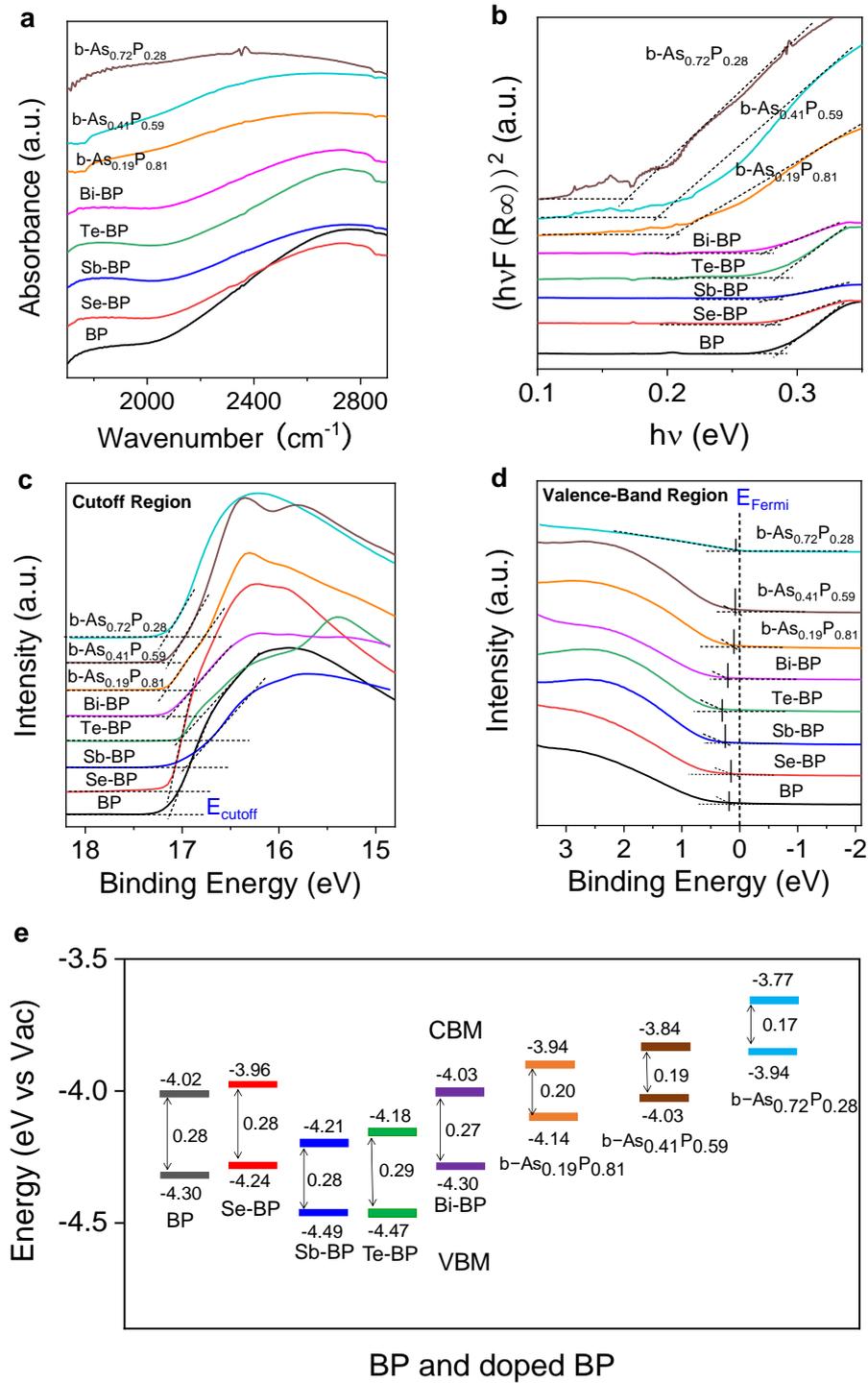

**Figure 4. Tuning of electronic properties of BP produced by doping.** (a) DRIFT absorbance spectroscopy, (b) hv-(hv*F(R∞))$^2$ curves; (c) Cutoff region of UPS spectra, (d) Valence-band region of UPS spectra. An offset has been added to the curves for better clarification for (a-d). (e) Energy band alignment of pristine and doped BP materials.

It is well known that BP is not stable in ambient environment, which significantly hinders its applications. To study the effect of doping on the ambient degradation of BP, we carried out AFM analyses on topographic features of the pristine BP flake, as well as Sb and Te-doped BP flakes as a function of exposure time (in lab light illumination, the humidity ranging from 70% to 91%, and the temperature ranging from 298 K to 303 K, **Figure 5** and **Figure S13, Supporting Information**). Similar to the reports by Wood et al.[47] and Yang et al.[48], hole-shape features are observed to appear on the surface of pristine BP flake, Sb and Te-doped BP flakes due to their reaction with adsorbed water and/or oxygen on the surface[49, 50]. With increased exposure time, the "hole" grow in size and diameter. The pristine BP flake exhibits faster growth of the "hole" and faster increase in diameter than the Sb and Te-doped BP flakes. **Figure 5a** shows an pristine BP flake of 25 nm in as-exfoliated form with ambient exposure up to 12 days. The "hole" evolves into large sizes with different shapes, some larger than 1 μm **(Figure S15, Supporting Information).** Severe corrosion of the pristine BP flake is clearly observable. A thinner pristine BP flake (25 nm in as-exfoliated thickness) obviously showed large "hole" structure after 16 days exposure due to corrosion (**Figure 5e** and **Figure S14a, Supporting Information**). In contrast, hole are uniformly distributed with much smaller sizes on a Sb-doped BP flake (28 nm thick in as-exfoliated flake) and Te-doped BP flake (31 nm thick in as-exfoliated flake), shown in **Figure 5b** and **Figure 5c**, with exposure up to 12 days. Surprisingly, with the further exposure to 16 days, the bubbles show little further change in size and shape. Even after 16 days of exposure, no severe corrosion is observable on the Sb and Te-doped BP flakes. The growth of hole on the flakes implies increase in surface roughness (see the height profiles in **Figure S14, Supporting Information**). With increased exposure time, roughness of the Sb and Te-doped BP flakes show much slower increase than that of the pristine BP flake (**Figure 5d**). Both pristine BP flake and Sb (Te)-doped BP flakes show a monotonic decrease in thickness with increased exposure time (**Figure** 5e). However, while the pristine BP flake experienced a thickness reduction by 32 % (8.0 nm, from 25 to 17 nm) after an exposure of 16 days, the Sb-doped BP flake experienced only 14.3 % reduction (4.0 nm, from 28 to 24 nm) in 16 days, and the Te-doped BP flakes experienced only 9.7 % reduction (3.0 nm, from 31 to 28 nm) in 16 days, as shown in **Figure 5e**.

As revealed in recent studies on the role of oxygen and water in ambient degradation of BP[49-52], the ambient degradation of BP is initiated by contact with oxygen rather than water, which leads to the first formation of $P_xO_y$ on the surface. In the presence of water, however, the produced $P_xO_y$ can be removed from the surface owing to its reaction with moisture to form phosphoric acid[53], and the BP beneath is exposed again to oxygen for continuous oxidation. Therefore, prevention of the initial reaction of BP with oxygen becomes crucial to inhibit the ambient degradation of BP. The energy band alignment in **Figure 4e** shows that, the VBM and CBM of pristine BP with respect to vacuum energy are -4.30 and -4.02 eV separately, which gives a band gap (**Figure 4b**) of 0.284 eV. With this sized band gap, the pristine BP can produce excitons under ambient light. Furthermore, the redox potential of $O_2/O_2^-$ is just located in the band gap (**Figure 5f**). Thus, the photo-generated electrons can transfer from the conduction band of BP to the $O_2$ on the surface and generate $O_2^-$, which would be apt to further react with the p-doped BP. While the Sb-doping and Te-doping downshift the CBM of BP significantly close to or even below the redox potential of $O_2/O_2^-$. In particular, for Sb-doped BP and Te-doped BP flakes, they cannot produce $O_2^-$ under ambient conditions at all as its CBM is too low (**Figure 5f**), and it is expected to be rather stable, consistent with our experimental results. Therefore, the bandgap-dependent exciton generation process combined with the CBM dependent charge-transfer process would make the pristine BP produce more $O_2^-$ than the Sb (or Te)-doped BP and eventually accelerate the degradation of BP observed in experiments.

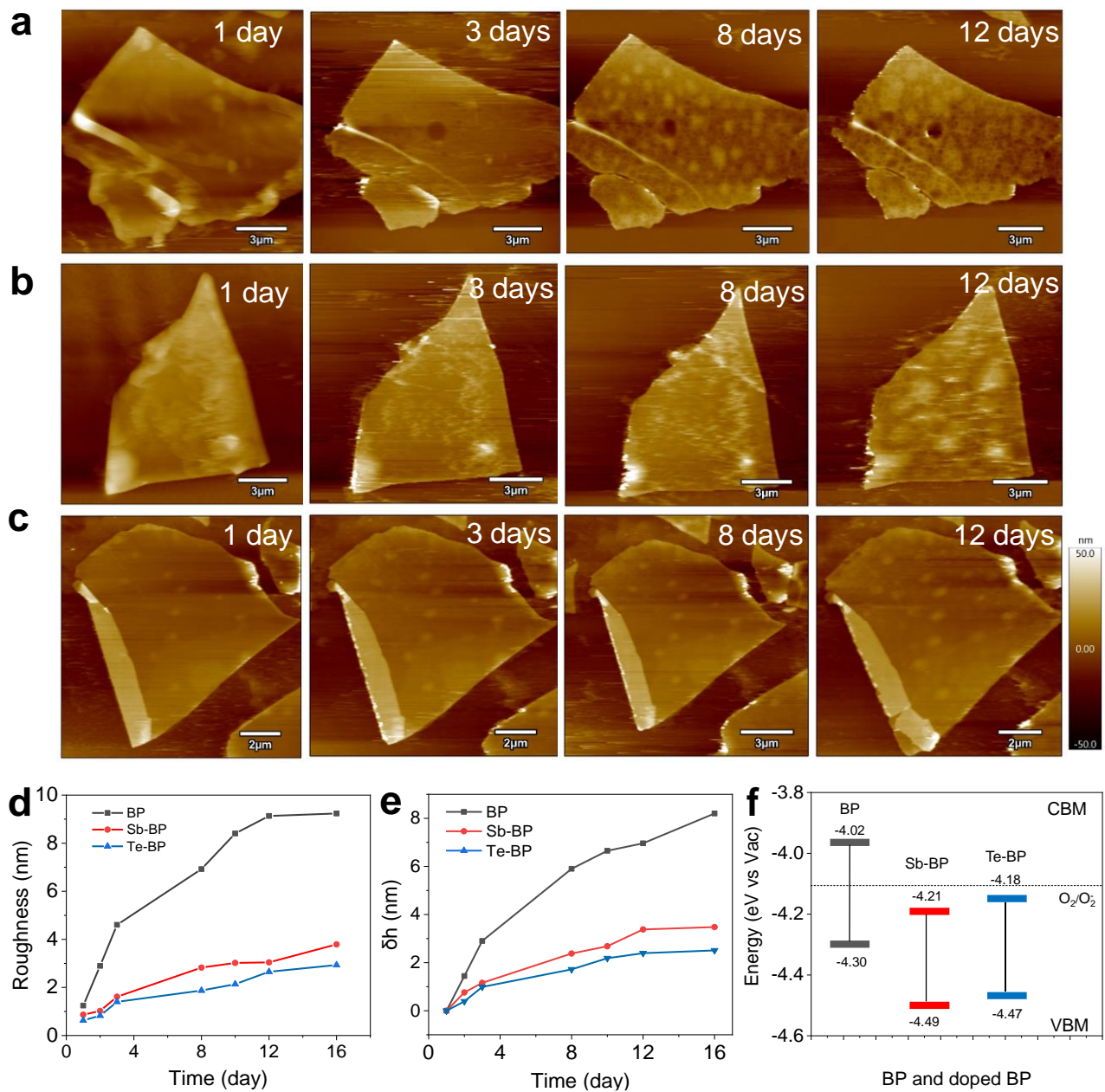

**Figure 5. AFM characterization of BP flakes and doped BP flakes upon ambient exposure.** (a) AFM images of a pristine BP flake with an as-exfoliated thickness of ≈25 nm, taken after ambient exposure for 1, 3, 8, and 12 days. (b) AFM images of a Sb-doped BP flake with an as-exfoliated thickness of ≈28 nm, taken after ambient exposure for 1, 3, 8, and 12 days. (c) AFM images of a Te-doped BP flake with an as-exfoliated thickness of ≈31 nm, taken after ambient exposure for 1, 3, 8, and 12 days. (d) Variation of surface roughness with increased exposure time. (e) Pristine BP, Sb-doped BP and Te-doped BP flakes thickness changes with exposure time. (f) The VBM and CBM of pristine BP and Sb, Te-doped BP with respect to vacuum energy. The dashed line identifies the redox potential of $O_2/O_2^-$.

In conclusion, we have developed a powerful SDT method using a uniform temperature to grow high-quality BP and doped BP with the highest growth yield up to 98%. In addition, uniform and controllable doping of various elements into BP was achieved with the same method, with the highest dopant concentrations reported so far. Structural and optical characterization show that the as-grown pristine and doped BP have high crystallinity. More importantly, we have also shown that doping BP effectively tunes its electronic structures including bandgap, work function, and energy band position. As a result, we have found that the air-stability of doped BP samples (Sb and Te-doped BP) improves compared with pristine BP due to the downshift of the CBM with doping. We envision that the doping of BP with tunable properties will significantly expand the uses of these materials in various areas.

**Supporting Information**

Supporting Information is available from the author.

**Acknowledgement**

We thank Changjiu Teng and Jie Ren for helpful discussions. We acknowledge support from the National Natural Science Foundation of China (No. 51722206), the Youth 1000-Talent Program of China, the National Key R&D Program (2018YFA0307200), Guangdong Innovative and Entrepreneurial Research Team Program (No. 2017ZT07C341), the Economic, Trade and Information Commission of Shenzhen Municipality for the "2017 Graphene Manufacturing Innovation Center Project" (No. 201901171523), and the Development and Reform Commission of Shenzhen Municipality for the development of the "Low-Dimensional Materials and Devices" discipline.

**Author contributions**

M.Q.L., S.M.F., and B.L. conceived the idea. M.Q.L. conducted most materials growth and characterization experiments. S.L.Z. performed PL and analysis under the supervision of F.W.. Y.H. and L.T. performed some of the XRD, XPS, AFM, and TEM characterization and analysis. J.M.L.

discussed the results and helped prepare the figures. B.L. supervised the project and directed the research. M.Q.L., S.M.F., Y.H., and B.L. interpreted the results and wrote the manuscript with feedback from the other authors.

**Conflict of Interest**

The authors declare no conflict of interest.